\documentclass[fleqn,useAMS]{mnras}

\usepackage{graphicx}
\usepackage{lscape}   

\title[Searching for giant planets in the outer Solar System]{Searching for giant planets in the outer Solar System with far-infrared all-sky surveys}

\author[Chris Sedgwick et al.]
{Chris Sedgwick and Stephen Serjeant
\\
School of Physical Sciences, The Open University, Milton Keynes MK7 6AA\\
}

\begin{document}

\date{Draft revised 5 July 2022}

\pagerange{\pageref{firstpage}--\pageref{lastpage}} \pubyear{2021}

\maketitle

\label{firstpage}

\begin{abstract}
We have explored a method for finding giant planets in the outer Solar System by detecting their thermal emission and proper motion between two far-infrared all-sky surveys separated by 23.4 years, taken with the {\it InfraRed Astronomical Satellite} (IRAS) and the {\it AKARI Space Telescope}. An upper distance limit of about 8,000 AU is given by both the sensitivities of these surveys and the distance at which proper motion becomes too small to be detected. This paper covers the region from 8,000 AU to 700 AU. We have used a series of filtering and SED-fitting algorithms to find candidate pairs, whose IRAS and {\it AKARI} flux measurements could together plausibly be fitted by a Planck thermal distribution for a likely planetary temperature. 
Theoretical  studies have placed various constraints on the likely existence of unknown planets in the outer solar system. The main observational constraint to date comes from a WISE study: an upper limit on an unknown planet's mass out into the Oort cloud. Our work confirms this result for our distance range, and provides additional observational constraints for lower distances and planetary masses, subject to the proviso that the planet is not confused with Galactic cirrus. 
We found 535 potential candidates with reasonable spectral energy distribution (SED) fits. Most would have masses close to or below that of Neptune ($\sim$0.05 Jupiter mass), and be located below 1,000 AU. However, examination of the infrared images of these candidates suggests that none is sufficiently compelling to warrant follow-up, since all  are located inside or close to cirrus clouds, which are most likely the source of the far-infrared flux.
\end{abstract}


\section{Introduction}\label{sec:intro}

The search for new planets in the solar system has attracted much work over the centuries. In general, this has been prompted by the need to explain observed perturbations in the orbits of known planets, and has involved trying to observe the planets from their reflected sunlight in the optical region of the spectrum. The discovery of Uranus in 1781, Neptune in 1846 and Pluto in 1930 resulted from such work. Much more recently, analysis of perturbations in the orbits of objects in the Kuiper Belt has suggested the likelihood of these being caused by a planet further out in the solar system (see review in Batygin et al.\ 2019).

Observing reflected sunlight is not the only method by which planets may be detected in the Solar System. Judging from the known giant planets, large planets are non-negligible sources of heat which may be detectible in the infrared. The peak of this radiation will depend on the temperature at the highest levels of the atmosphere of the planet, and will most likely be in the far-infrared (see Table \ref{table:planets2}). Since reflected sunlight will fall by r$^{-4}$ with distance r from the Sun, detection of emitted thermal radiation (which will fall by r$^{-2}$) becomes a much more likely method of detecting planets in the Solar System at greater distances from the Sun. Detection in this way using a planet's proper motion has only recently become possible, with two all-sky far-infrared catalogues now available, taken some 23 years apart.

The most distant giant planet currently known, Neptune, orbits the sun at a semi-major axis of $\sim30$ AU. The Kuiper Belt covers the region 30 AU to 55 AU and has three known dwarf planets (including Pluto) and numerous smaller objects (Kuiper Belt Objects, KBOs). A Neptune-sized planet at $\sim700$ AU was recently predicted to explain the eccentric orbits of several such objects (\textquoteleft Planet Nine\textquoteright ; Batygin \& Brown 2016). Our longer term objective is to cover this region, but this initial paper only reaches its edge, covering from 700 AU out to almost the beginning of the Oort Cloud, which is generally thought to extend from $\sim$ 10,000 AU to 100,000 AU. Socas-Navarro (2022) used anomalies in the velocity and trajectory of a meteor to support the existence of a ninth planet.
Another recent paper reported an anomaly in the orbits of outer Oort Cloud comets, which may suggest a  planet of 1$-$4 Jupiter masses (M$_{\rm J}$) at $\sim10,000-30,000$~AU (Matese \& Whitmire 2011): however, our work would not detect a planet as distant as the Oort Cloud. 

Individual all-sky infrared catalogues have previously been examined for evidence of new planets using parallax from orbital extremes of the space telescope. The IRAS Point Source Catalogue (PSC) was searched for evidence of \textquoteleft Planet X\textquoteright~shown by a slowly moving blackbody (e.g.\ Houck et al.\ 1985) but the sources identified turned out not to be planets (see also review in Beichman 1987). The rejected sources catalogue (PSCR) in the IRAS survey was examined for moving sources (Davies et al.\ 1984) and three asteroids and six comets were found. A possible detection of an unknown object as large as Jupiter in the solar system was made by IRAS in December 1983, but further analysis showed it to be a known comet (Rowan-Robinson 2013). More recently, Luhman (2014) searched for planets by parallax in WISE data, but without success, reaching distances for potential Saturn-mass and Jupiter-mass planets of 28,000 AU and 82,000 AU respectively. There was a recent tentative detection of a distant source in the Solar System by ALMA, although this paper has been withdrawn pending further analysis and new data (Liseau et al.\ 2015). Using IRAS data, a candidate for a planet of 3$-$5 Earth masses (M$_{\rm E}$) at 225$\pm$15 AU has recently been identified (Rowan-Robinson 2021).

There have also been previous studies looking for evidence of new planets using proper motion between surveys. A recent all-sky survey looked for objects by their proper motion using optical catalogues, identifying eight already-known objects in the Kuiper Belt at 31 AU to 68 AU (Brown et al.\ 2015). Luhman \& Sheppard (2014) looked for high proper motion objects in the near- and mid-infrared by comparing 2MASS and WISE data, but focused on nearby stars rather than planets. Our work takes a similar approach in the far-infrared, aiming to identify objects by their proper motion in the 23.4 year period between two all-sky far-infrared surveys (IRAS and {\it AKARI}) to try to identify a distant planet in the Solar System. 

\begin{figure}
\begin{center} 
\resizebox{3.3in}{3.3in}{\includegraphics{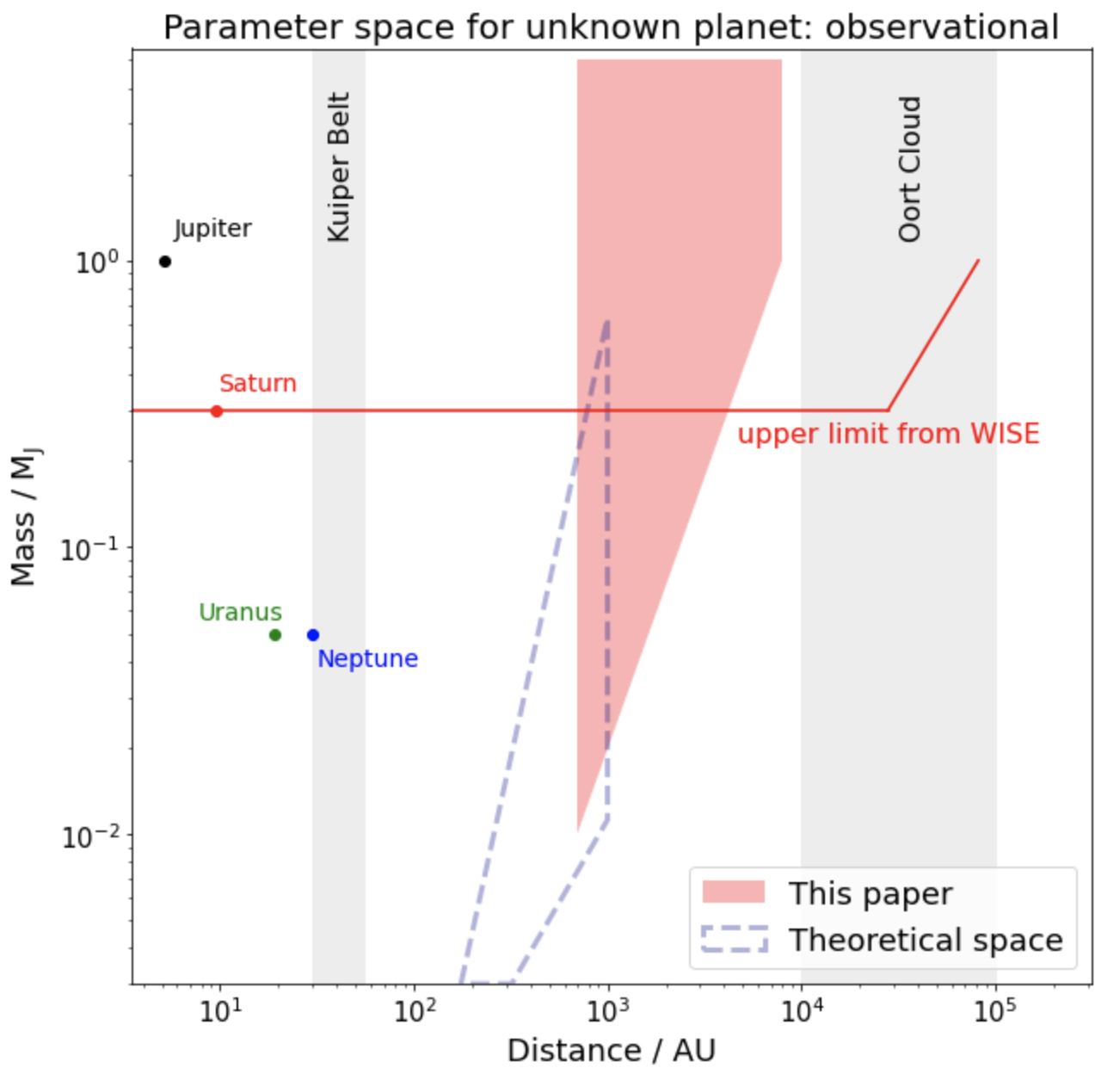}} 
\resizebox{3.3in}{0.22in}{\includegraphics{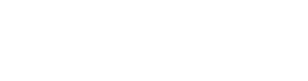}}
\resizebox{3.3in}{3.3in}{\includegraphics{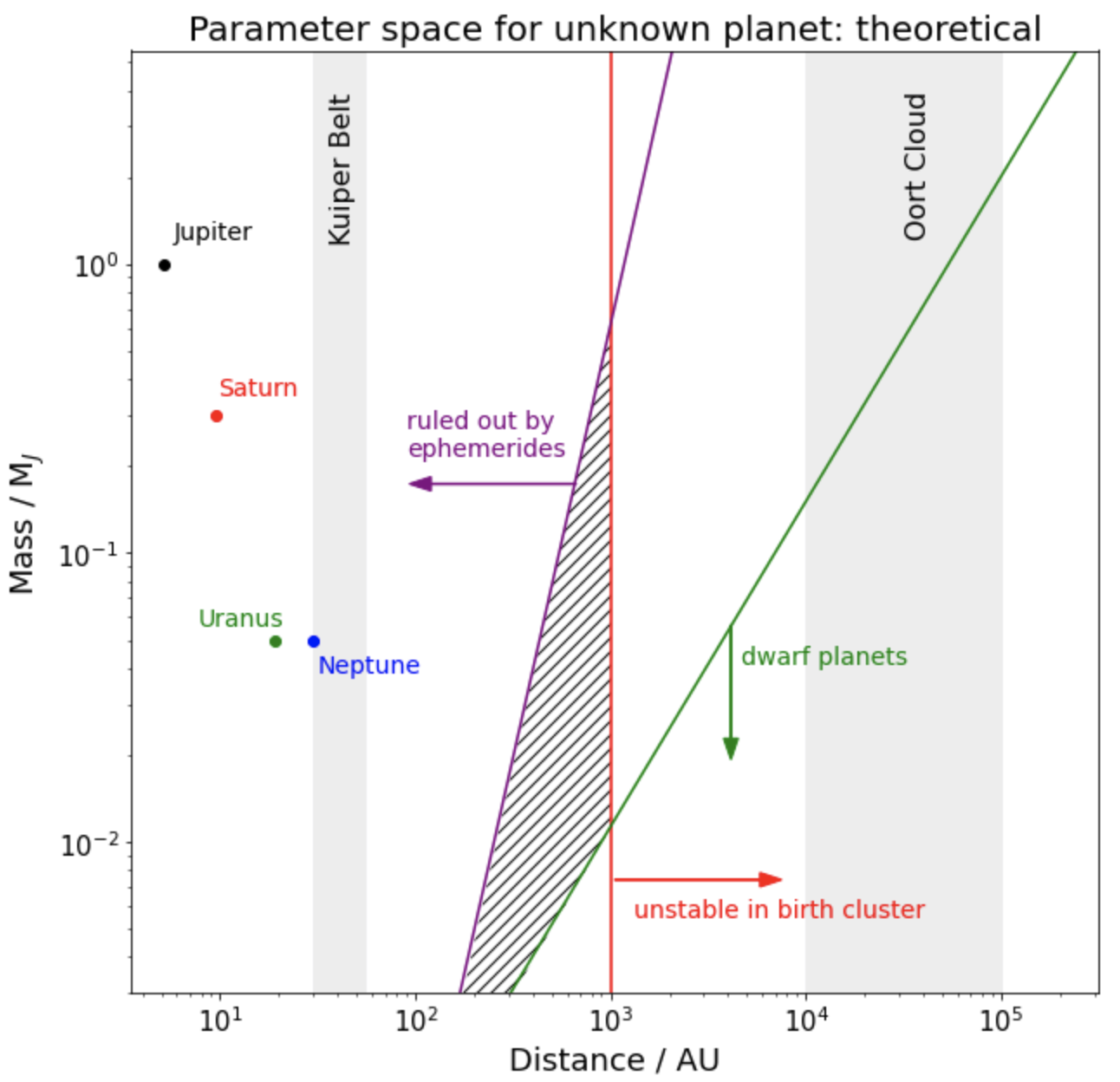}} 
\caption{Top: observational constraints on an unknown planet, from WISE (red line; Luhman 2014) and from this project to date (red polygon). The outline blue polygon shows the theoretical constraints shown below. Bottom: theoretical constraints on an unknown planet: the purple line indicates the parameter space ruled out by perturbations in known giant planets;  the red line indicates the theoretical limits for the survival of a planet from the solar birth cluster; the green line shows the limits of the definition of a planet which has cleared out its orbit over the age of the solar system (see Batygin et al.\ 2019 and references therein). The black hatched area is the theoretical space for the planet resulting from these constraints.}\label{fig:parameter_space}
\end{center}
\end{figure}

The review article by Batygin et al.\ (2019) combines various theoretical studies with the WISE results to suggest that the parameter space for possible undiscovered planets in the Solar System lies between 100 $-$ 1,000 AU, and between 1$-$100 Earth masses ($\sim 0.003 - 0.3$ M$_{\rm J}$) (Figure 5 of that paper). We have covered part of this space (700 AU to 1,000 AU) in this paper (illustrated in Figure \ref{fig:parameter_space}). We also covered the region out to 8,000 AU, since the data are available and can provide confirmation of part of the results of the WISE study.

The IRAS and {\it AKARI} all-sky infrared surveys are described in Section~\ref{sec:surveys}. The detectability in these surveys of a potential planet's flux, parallax and proper motion is discussed in Section~\ref{sec:detectability}. The algorithms used to narrow the search for potential planets by rejecting unlikely candidate pairs are described in Section~\ref{sec:algorithms}. The resulting shortlist is considered in Section \ref{sec:shortlist}, and the results are discussed  in Section~\ref{sec:results}.

\section{All-sky far-infrared surveys}\label{sec:surveys}

\begin{table}
\caption{Orbital data on presently known giant planets. Sources: Guillot \& Gautier (2015); McBride \& Gilmour (2004).}\label{table:planets2}
\begin{center}
\begin{tabular}{|lrrrr}
\hline
                                                          &  Jupiter           & Saturn               & Uranus             & Neptune \\
 \hline
Inclination to ecliptic                        &  1.3$\degr$    & 2.5$\degr$    & 0.8$\degr$    & 1.8$\degr$     \\      
 Orbital period (yr)                           &   11.86                &  29.42             & 83.75              & 163.7  \\
Semi-major axis (AU)                      &  5.20                  & 9.54                & 19.19              & 30.07   \\
Equatorial radius (km)                     & 71,400               & 60,300            & 25,600            & 25,200   \\
Mass (10$^{24}$kg)                        & 1,900                   & 568                 & 87                   & 102     \\
Mass (M$_{\rm J}$)                        &  1.00                   & 0.30               & 0.05                & 0.05    \\
Eff. temperature (K)                        & 124                     & 95                  & 59                   & 59     \\
Blackbody peak  ($\mu$m)             &  23                      & 31                  & 49                   & 49 \\
\hline
\end{tabular}     
\end{center}
\end{table}

\begin{table}
\caption{Central wavelengths of IRAS and AKARI instruments with equivalent peak blackbody temperature.}\label{table:planets1}
\begin{center}
\begin{tabular}{|lrr}
\hline
                                                          & Central     &  Equivalent    \\
                                                        &  Wavelength     &   Temperature    \\
                                                              &       ($\mu$m)    &      (K)              \\
 \hline
 IRAS                                                      &     12     &   241      \\
IRAS                                                      &      25  &    116       \\
IRAS                                                 &      60       &    48            \\
IRAS                                                      &       100  &      29      \\\\
 {\it AKARI} - FIS N60                               & 65   &    45   \\
 {\it AKARI} - FIS Wide-S                           &  90   &  32      \\
 {\it AKARI} - FIS Wide-L                      &  140      & 21          \\
 {\it AKARI} - FIS N160                         &   160    & 18        \\
 \hline                                
\end{tabular}     
\end{center}
\end{table}

\begin{table*}
\caption{Numbers of sources in the IRAS and {\it AKARI} catalogues including the IRAS Rejects catalogues. Depth and resolution are shown for 100 $\mu$m for IRAS and at 90 $\mu$m for {\it AKARI}. Identified sources are those identified in the IRAS published catalogues. *The reliability of  IRAS-FSCR is likely to be as low as 10\%. }\label{table:catalogues}
\begin{center}
\begin{tabular}{|lllrrrrrr|}
\hline                              
Catalogue                &Wavelengths      & Date                             & Depth        & Resolution                         & Positional            & Total               & Identified  & Unidentified   \\
                                    &  ($\mu$m)         & (mid-survey)                &  (Jy)           &  PSF FWHM                     & accuracy            & sources        & sources   & sources \\
 \hline
IRAS-PSC                 & 12, 25, 60, 100   & Jun 1983                 &  0.6           & 60$\arcsec$                         & 12$\arcsec$     &  245,889       &  85,047 & 160,842   \\
IRAS PSCR             &   12, 25, 60, 100  &  Jun 1983                &                  &                                                 &                       &  372,753   &   51,000    &  321,753  \\
IRAS-FSC                 &   12, 25, 60, 100  &  Jun 1983                &    0.2        &                                                  &  15$\arcsec$    & 173,044  & 115,273     &  57,771    \\
IRAS-FSCR*            &   12, 25, 60, 100  &  Jun 1983                &                  &                                                  &                           &  593,516  &  211,144  & 382,372  \\\\
AKARI-BSCv2            & 65, 90, 140, 160  & Nov 2006                &  0.44        &  39$\arcsec$                         & 8$\arcsec$       & 918,056           &                     &      \\

\hline
\end{tabular}
\end{center}
\end{table*}

The catalogues from two all-sky surveys which cover far-infrared wavelengths, by IRAS and {\it AKARI}, have been used to identify planet candidates. Both surveys cover wavelengths near the flux peaks for known giant planets in the solar system, with {\it AKARI} supplying coverage of the shallow Rayleigh-Jeans tail of an assumed thermal distribution (see Tables \ref{table:planets2} and \ref{table:planets1}).

\subsection{IRAS} 
Firstly, the all-sky survey by the Infrared Astronomical Satellite (IRAS; Neugebauer et al.\ 1984) in 1983 was taken at wavelength bands centred on 12 $\mu$m, 25 $\mu$m, 60 $\mu$m and 100~$\mu$m and generated two large catalogues: the point source catalogue (IRAS-PSC) of some 246,000 sources and the faint source catalogue (IRAS-FSC) which was generated by stacking the multiple scans which yielded some 173,000 sources of which 58\% are not in the IRAS-PSC catalogue. The IRAS-PSC reached a depth of 0.6 Jy, whereas the IRAS-FSC reached $\sim$0.2 Jy at 60 $\mu$m. There is also a catalogue of rejects for each of these catalogues. Since one of the reasons for rejection could be failure to confirm the location of a source in multiple scans, these are also relevant, perhaps crucial, to our search (see Table \ref{table:catalogues} for further details). 

This paper concentrates on far-infrared observations. We have not used the lowest IRAS wavelength (centred on 12~$\mu$m) in the blackbody SED-fitting algorithms, since the known gas giants have higher mid-infrared flux than that given by the sharp Wien tail fall-off of single-temperature blackbody SEDs. This is discussed further below (Section \ref{sec:fitting}).

\subsection{AKARI} 
Secondly, the {\it AKARI Space Telescope} (Murakami et al.\ 2007) all-sky survey of 2006-07 generated a bright source catalogue ({\it AKARI}-BSC; Yamamura et al.\ 2010).  Observations at 65~$\mu$m, 90 $\mu$m, 140 $\mu$m and 160 $\mu$m  were made with the Far-Infrared Surveyor (FIS; Kawada et al.\ 2017). Version 2 of this catalogue was released in 2016 (Yamamura et al.\ 2017) and contains 918,056 sources. The 90~$\mu$m observations reached a depth of 0.44 Jy and had a positional accuracy of $\sim8\arcsec$, with multiple scans of most of the sky. . We have used both the main catalogue (501,444 sources) and the supplemental catalogue (416,612 sources) for the same reason that we used both the main and the reject catalogues from IRAS.

\subsection{Mid-infrared all-sky catalogues: WISE and AKARI} 
There is a mid-infrared all-sky catalogue from the Wide-field Infrared Survey Explorer (WISE; Wright et al.\ 2010) in January-August 2010, covering four bands, at 3.4 $\mu$m, 4.6~$\mu$m, 12 $\mu$m and 22 $\mu$m (referred to as W1 to W4 respectively). The highest of these wavelength bands, the WISE 22 $\mu$m band is borderline far-infrared, and is close in wavelength to the lowest IRAS band at 25 $\mu$m and can be used (after extrapolating the position for a further $\sim$4 years) to check for possible confirmation of plausible candidate planets.

Observations for the first two WISE bands (W1 and W2) extended over some six years, up to 2016. However, observations at the two higher wavelength bands (W3 and W4) lasted for just 7 and 9 months respectively. Evidence of parallax (and, given the timescale) proper motion can be searched for using the W1 and W2 bands, and a number of studies have already been made of this data (e.g.\ Luhman 2014). If we find matches in the band 4 data to our candidates, we could potentially use these data to look for parallax in the WISE data.

{\it AKARI}-IRC also provides all-sky data in the mid-infrared (at 9 $\mu$m and 18 $\mu$m). However, there is no clear method of using these data in matching the far-infrared data discussed in the paper, since they would not relate to the fitted thermal distribution.

\subsection{Submillimetre all-sky catalogue: Planck}
The Planck all-sky survey (Planck-Collaboration 2011) covered submillimetre to millimetre wavelengths down to 350~$\mu$m. However, its depth at 350~$\mu$m was 0.7 Jy, not deep enough to detect emission from the Rayleigh-Jeans tail of giant planet SEDs for masses considered here.

\section{Detectability of a planet's thermal emission by IRAS and {\it AKARI}-BSC}\label{sec:detectability}

The upper limit for the distance at which we may detect a planet is set by the sensitivity of the IRAS and {\it AKARI} instruments in terms of (a) the flux observable and (b)  the proper motion detectable between the two surveys. We discuss each of these in turn below, together with the parallax detectable within each survey. We have made the simplifying assumptions of a circular orbit and a flux emission similar to that of the known giant planets for potential planets with masses below that of Jupiter. We have used the model predictions given in Burrows et al.\ (1997) to extrapolate temperature and radius for potential planets with masses greater than that of Jupiter. The calculations underlying the discussion below are shown in Table~\ref{table:proper_motion}. 

An inner limit is effectively the point where the perturbations caused by a planet would likely already have been observed, particularly by objects in the Kuiper Belt. We have taken 700 AU as the inner limit in this paper; we plan to extend inwards in future work.

\subsection{Detectability of flux}
 The infrared flux received from the giant planets exceeds that expected from the planets' reflection of the Sun's light, and is due to heat generated internally, most likely from gravitational collapse over a long time scale. This excess thermal radiation is likely to peak in the far infrared (see Table \ref{table:planets2}). The four known giant planets also reflect some infrared radiation from the Sun, but the flux received from this would be comparatively insignificant from planets further from the Sun.
 
Table \ref{table:catalogues} shows the detection limits for the far-infrared surveys. {\it AKARI} can detect down to about 0.44 Jy (slightly lower than the 0.55 Jy in Version 1 of the catalogue), and IRAS to 0.5 Jy for PSC and 0.2 Jy for FSC. The flux detection limit for detecting sources matched between the two catalogues is therefore $\sim$ 0.5 Jy.
  
Table \ref{table:proper_motion} shows the flux receivable from a potential planet with mass similar to that of Uranus or Neptune (0.05 M$_{\rm J}$), of Saturn (0.3 M$_{\rm J}$) and of Jupiter, at various distances. We would not detect a Jupiter-mass planet much beyond about 6,000 AU with either IRAS or {\it AKARI}. The limit for a Saturn-mass planet (0.3 M$_{\rm J}$) is $\sim4,000$ AU; for a Neptune-mass planet it is $\sim$ 800 AU. On the other hand, a planet of 5~M$_{\rm J}$ could be detected out to about 10,000 AU.
 
It is worth considering the evidence from exoplanets on planet sizes (although there will be selection effects). Three-quarters of exoplanets being discovered by the {\it Kepler} mission are smaller than Neptune, and only 2\% are Jupiter-sized or above (Borucki et al.\ 2011; Howard et al.\ 2013). However, planets at the large distances from their stars considered in this paper (over 700 AU) are not likely to be detected in the work on other solar systems: virtually all exoplanets detected so far are within 100 AU (see review in Winn \& Fabrycky 2015). However, brown dwarfs have been found for a few nearly stars at distances $>1,000$~AU (e.g.\ Luhman et al.\ 2011).


\begin{table*}
\caption{Estimates of flux, parallax and proper motion over the 23 years between IRAS and {\it AKARI}. The table shows a Saturn-mass planet (=0.3M$_{\rm J}$) drops below detectability at around 4,000 AU, and a Uranus-mass planet is potentially detectable out to $\sim$1,000 AU. Estimates of proper motion assume circular orbit and Kepler's third law.}\label{table:proper_motion}
\begin{center}
\begin{tabular}{lrrrrrrrrrrrrrr|}
\hline   
                  &  \multicolumn{8}{c|}{ ---------------------- Estimates of planetary flux  ---------------------- }                                                                     &&   \multicolumn{3}{c|}{ Estimates of proper motion }  && Parallax \\  
                  &  \multicolumn{2}{|c|}{-- Uranus mass --}  &  &  \multicolumn{2}{|c|}{-- Saturn mass --}  & &  \multicolumn{2}{|c|}{-- Jupiter mass --}     &    &                &   &  &   \\ 
Distance            &60 $\mu$m   & 90 $\mu$m  & & 60 $\mu$m                & 90 $\mu$m       &&  60 $\mu$m       & 90 $\mu$m                              &   & Orbit               &  \multicolumn{2}{c|}{Proper Motion}   &  & \\
 (AU)    & S$_{\lambda}$/Jy  & S$_{\lambda}$/Jy &  & S$_{\lambda}$/Jy  & S$_{\lambda}$/Jy      &    & S$_{\lambda}$/Jy  &S$_{\lambda}$/Jy        & &  /years        &        p.a.                   &      23.4 yrs      &  &    \\  
                  
 \hline
 
500                    & 1.1~              & 1.4~               &  &  32~~           & 25~~               &   &   66~~         & 48~~                  &  &   1.1$\times10^4$     & 116\arcsec     & 45.2\arcmin      &  &  6.9\arcmin \\
700                   &  0.6~              & 0.7~               &&  17~~            & 13~~                &&     34~~          & 24~~                  &&    1.9$\times10^4$      &  70\arcsec      & 27.3\arcmin      &&   4.9\arcmin \\
800                    &   0.4~            & 0.5~              &  &  13~~           &  9.8                  &   &   26~~         &    19~~               &  &    2.3$\times10^4$    &  57\arcsec    &   22.4\arcmin     &  &  4.3\arcmin       \\
1,000                 & 0.3~               & 0.3~              &   &   8.1            &  6.3                  &   &   17~~         &  12~~                 &  &   3.2$\times10^4$     & 41\arcsec      & 16.0\arcmin       &  & 3.4\arcmin  \\
2,000                 & 0.07               & 0.09              &   & 2.0               & 1.6                  &  &    4.1            &  3.0                     & &   8.9$\times10^4$     & 15\arcsec       & 5.7\arcmin         &  & 1.7\arcmin \\
3,000                 & 0.03               & 0.04              &  &  0.9               &  0.7                &  &    1.8             & 1.3                      &  &  1.6$\times10^5$      &  7.9\arcsec    &  3.1\arcmin        &  & 1.1\arcmin \\
4,000                 & 0.02               & 0.02             &   &   0.5             & 0.4                 &  &    1.0             &  0.7                     &  &   2.5$\times10^5$     &  5.1\arcsec     & 2.0\arcmin        &  & 51.6\arcsec \\    
6,000                 &   0.01              & 0.01             &   &  0.2              & 0.2                 &   &    0.5            &  0.3                     & &     4.6$\times10^5$     &  2.8\arcsec    & 1.1\arcmin        &  & 34.4\arcsec \\    

\hline
\end{tabular}
\end{center}
\end{table*}

\subsection{Detectability of proper motion}
There was a period of some 23.4 years between the observations by IRAS and by {\it AKARI}, taking mid-points of the survey periods in each case. Table \ref{table:proper_motion} shows the estimated proper motion of a planet over this period at various distances. As shown in Table \ref{table:catalogues}, the angular resolution of IRAS and {\it AKARI} was $\sim60$\arcsec~and 39\arcsec~respectively. The IRAS resolution suggests that a source over $\sim$6,000 AU may not be resolved. The filters in our algorithms to identify candidates take sources within 30$\arcsec$ to be matches between the catalogues, and excludes them. At 700 AU, the lower limit of our survey in this paper, the proper motion over 23.4 years would be 27.3\arcmin. The positional uncertainties of both surveys (see Table \ref{table:catalogues}) were smaller than the proper motion out to $\sim8,000$ AU (see Table~\ref{table:proper_motion} and Figure~\ref{fig:limits}).

\subsection{Detectability of parallax}

\begin{figure}
\begin{center} 
\resizebox{3.03in}{!}{\includegraphics{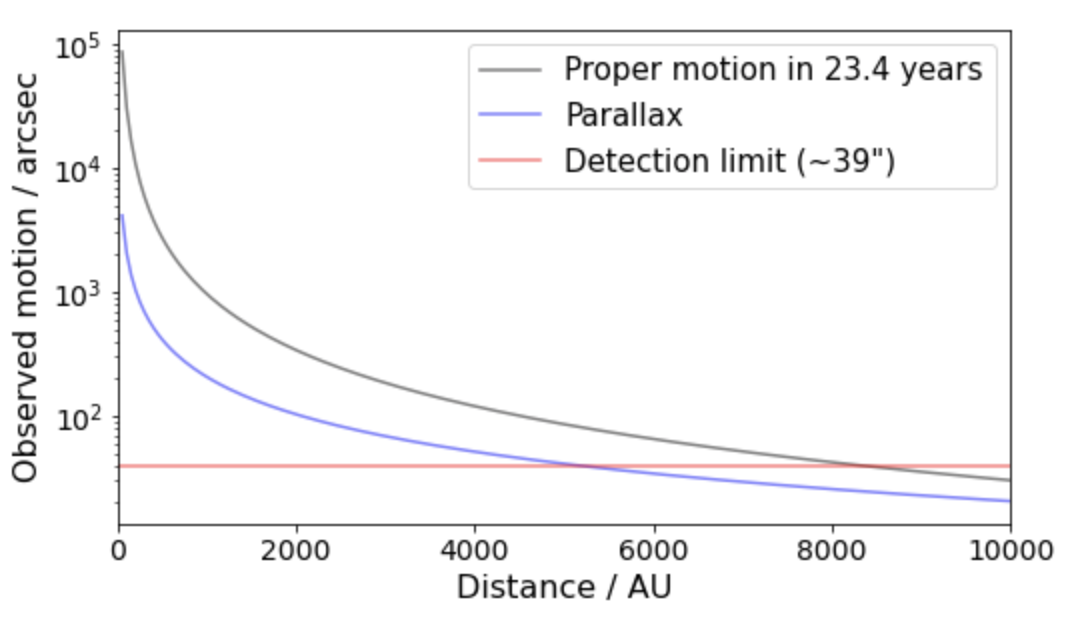}} 
\caption{Proper motion and parallax observed angular motion compared to an estimated detection limit of 39\arcsec.  }\label{fig:limits}
\end{center}
\end{figure}

Parallax for the relevant distances is also shown in Table \ref{table:proper_motion} (last column).
It is not detectable with the beam sizes of IRAS and {\it AKARI} for objects at distances greater than $\sim5,000$ AU (see also Figure \ref{fig:limits}). At closer distances, it provides an alternative method of detecting sources; this was unsuccessful in previous studies with IRAS (see Section \ref{sec:intro}) and multiple observation data are not (yet) available for {\it AKARI}. Parallax may cause problems for our detection by proper motion at lower distances: we have taken median values in SCANPI (see Section \ref{sec:scanpi}) partly for this reason. We discuss the evidence of parallax in IRAS data in Section \ref{sec:scanpi} below.

\subsection{Assumptions and approximations}

We have assumed orbits with low eccentricity. If this is not true, our estimates of distance and mass will be affected, but detection should still occur. It is quite possible that we should expect eccentric orbits in view of the infrequency of comets and asteroids knocked in the direction of Earth (and the eccentricity of Kuiper Belt Objects like Pluto is high).

The four presently known giant planets lie close to the ecliptic plane (see Table \ref{table:planets2}), the highest inclination being Saturn at only 2.49$\degr$. New giant planets close to the ecliptic plane, however, are likely to have detectible effects on already-known objects, so probably planets at higher orbital inclinations to the ecliptic are more likely to remain to be discovered. This is increasingly likely as AU is reduced.
We have ignored the effect of the Earth's orbit: at the separations being considered, 2 AU would have only a tiny effect.

\subsection{Asteroids and comets}

Asteroids and comets are bright sources in the infrared. However, asteroids are fast-moving objects, so our candidates would not be asteroid detections. The distance of the major Asteroid Belt (2.1 AU$-$3.3 AU) means that their proper motion would be 10\arcmin$-$20\arcmin~per day (and their orbits would be 3$-$6 years): our study has produced candidates with proper motion which is a tiny fraction of this. Also, the emission from asteroids can vary significantly over short timescales, and the sort of matching we have done by combining fluxes from two surveys would not detect sources with varying flux. Furthermore, where multiple scans, even a day apart, have detected an object, this would rule out identification with an asteroid in the Asteroid Belt. There are also asteroids further out, but these are still far closer, and on far faster orbits, than objects considered in this paper.
 
 However, it is possible that some of the individual IRAS scans in our candidate pairs may be asteroids. Matching to asteroids and comets can be done using the NEOChecker at the IAU Minor Planet Center\footnote{http://www.minorplanetcenter.net/iau/mpc.html.}, which currently has over a million near-earth objects including asteroids, comets and other objects. Using the precise observation times given by SCANPI, we can check the scans of any plausible planet candidates.

\section{The algorithms}\label{sec:algorithms}

\subsection {Identification of possible pairs}

Our essential approach has been to find pairs of sources across the two catalogues, 7 of whose 8 flux measurements (excluding the 12 $\mu$m flux in IRAS, as discussed in Section \ref{sec:fitting}) matched well enough to approximate a Planck blackbody distribution at a plausible planetary mass and temperature. So the objective of these algorithms is to identify candidate sources which have a broad approximation to a Planck distribution, for further investigation. This was done in successive proper motion ranges equivalent to objects at specific distance ranges from the observer in AU (see Table~\ref{table:planets4}).

The first step was to remove all sources which matched between the {\it AKARI}-BSC and IRAS catalogues, and this was done for each of the four IRAS catalogues in turn. Sources were taken as matched if they were within 30$\arcsec$ in each pair of catalogues; this cutoff was based on the resolution and positional accuracy of the catalogues (see Table~\ref{table:catalogues})\footnote{ The {\it AKARI} team also  identified sources back to IRAS using a larger matching radius of 100$\arcsec$, but their identifications are not included in the {\it AKARI} catalogue and have not been used here.}.

The next step was to remove already known sources. We excluded sources already identified to another catalogue by the IRAS team, as indicated by a NID value in the IRAS catalogue. The IRAS team matched to 41 catalogues including asteroids and comets. The matching radius used varied with the positional accuracy of the various catalogues, and was typically 90\arcsec~(details are given in the IRAS Explanatory Supplement\footnote{The IRAS Explanatory Supplement is available at http://irsa.ipac.caltech.edu/IRASdocs/exp.sup/.}). Table \ref{table:catalogues} shows the number of sources identified by IRAS and therefore excluded from our search. Previous identification of already-known sources was not available for the {\it AKARI} catalogue when this work was done. Sources identified more recently need to be checked, but the time spent checking some 2.3 million sources to various catalogues would seem to be greater than that saved in running the algorithms for a slightly reduced number of sources, so we decided to do this later when we have a shortlist of potential candidates.

We then identified all possible pairs within a selected AU range between the {\it AKARI}-BSC and each of the four IRAS catalogues in turn, starting with 6,000$-$8,000 AU and working inwards in increasingly narrow bands as the number of possible pairs increased. The result at this stage was four databases of candidate pairs (one for each IRAS catalogue). The relation between the proper motion and distance was calculated assuming Keplerian orbits.

Looking at all the sources in the two catalogues to check broad compatibility, we found that although the  {\it AKARI} 65~$\mu$m flux was broadly consistent with the IRAS 60 $\mu$m flux, the {\it AKARI} 90 $\mu$m flux was strongly inconsistent with the IRAS 100 $\mu$m flux in all four IRAS catalogues. This inconsistency is already known from earlier studies (Serjeant \& Hatzimangalou 2009; Hereaudeau et al.\ 2004; Jeong et al.\ 2007). Rowan-Robinson \& Wang (2017) showed the need for an aperture correction to the AKARI 90 $\mu$m fluxes. We have accounted for this in a crude fashion across the catalogue as a whole, by using a fixed 1/0.7 correction for these fluxes.  


\subsection{Use of SCANPI data for IRAS fluxes}\label{sec:scanpi}


Most of the candidate sources found from the procedures described below were from the IRAS PSCR (i.e. the IRAS point source catalogue {\bf reject} catalogue). The problem this caused was that the fluxes in the sources matched to this catalogue were virtually all upper limits. To obtain better estimates of the fluxes and errors in the IRAS data, we therefore decided to use the Scan Processing and Integration tool (SCANPI; Helou et al.\ 1988), a revised version of which was made available in 2007. These values were often significantly different to the upper limits shown in the reject catalogue and are considered to be much improved data. The sensitivity gain in SCANPI\footnote{SCANPI User's Guide, September 2007, available at http:// irsa.ipac.caltech.edu/applications/Scanpi/docs/overview.html.} is comparable to that obtained with the IRAS Faint Source Catalogue.

SCANPI offers up to four estimates of a source's flux, all given in Janskys: (a) the peak flux, (b) the average flux density between zero-flux crossings, (c) the average flux density between fixed points from the target, with defaults set to $\pm$2$\arcmin$ for 12 $\mu$m and 25 $\mu$m, $\pm$2.5$\arcmin$ for 60 $\mu$m and $\pm$4$\arcmin$ for 100 $\mu$m, and finally (d) the peak flux density of the best-fitting point source template. Not all estimates were available for all sources. Following the guidance in the SCANPI User's Guide, we took the flux values in the following order of preference: the template amplitude; if not available, the flux density between the default points; failing that, the peak flux. The flux density between zero-crossings was not relevant since the continuum shown in the SCANPI reduction was not usually close to zero. If none of these three options was available, we stayed with the upper limit from the reject catalogue.

In addition, SCANPI offered four averages from the multiple scans (typically between 8 and 20 scans) for each source: straight mean, noise-weighted mean, median and mean with noisy detectors half-weighted. These were usually fairly close together. We took the median as the most appropriate average for our purposes. 

The noise measurement (root-mean-square deviation of the residuals after the baseline subtraction) for the coadds uses a Local Background Fitting Range (we took the default of 30\arcmin) excluding the central Source Exclusion Range (we took the defaults of 2.0\arcmin, 2.0\arcmin, 4.0\arcmin~and 6.0\arcmin~for the four bands respectively).

Details of individual scans are available with SCANPI. Where scans were taken with a long time interval between them (ideally 6 months) this gives an opportunity to check if parallax is observed, particularly for the closer objects (at 700 AU, parallax is 4.9\arcmin). The SCANPI data provides the date and time of each scan (in Universal Time). The UT for each scan can be converted to Julian Day Number, and the Julian Day Number of the IRAS launch subtracted to give the number of days from launch. SCANPI also provides the in-scan deviation of the signal peak from the user-specified target position (a parameter called \textquoteleft MISS'). The location of the peak can be taken to be the centre of the best-fitting template. The MISS parameter can then be plotted against the days from launch for each shortlist source.\footnote{Data from individual scans by {\it AKARI} are not yet available.}

\subsection{Filters based on position: emission from galactic cirrus}

\begin{figure}
\begin{center} 
\resizebox{3.0in}{!}{\includegraphics{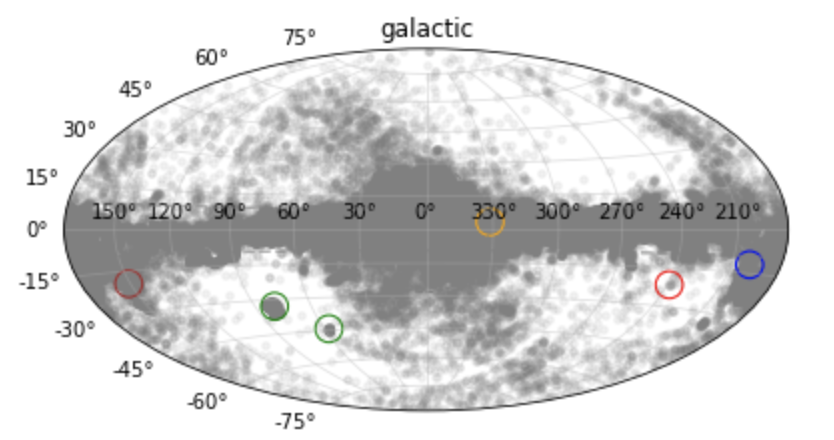}}  
\caption{All candidate pairs before filtering in the range 990 -1000 AU in galactic coordinates, showing preponderance of candidates close to the galactic plane, and close to the galactic centre, due to galactic cirrus. A similar pattern in the galactic plane and central bulge at mid- and far-infrared wavelengths was mapped by COBE/DIRBE (Bernard et al.\ 1994). The magenta circle shows the position of Andromeda; the green circles show the LMC and SMC; blue is the California nebula, brown is Orion and orange the Gould Strip. }\label{fig:galactic}
\end{center}
\end{figure}

Galactic cirrus is a significant problem when trying to identify planets at far-infrared wavelengths. A significant number of the pairs of candidates before filters were applied proved to be close to the galactic plane (see Figure \ref{fig:galactic}). A study of the dust emission of Galactic cirrus from COBE / DIRBE observations showed a concentration at low Galactic latitudes, with a broader extension near the Galactic bulge (Bernard et al.\ 1994). Therefore, as a first filter, we excluded sources with galactic latitude $|b|<10$\degr~and sources within a radius of 27.5\degr~of the Galactic centre.

These filters removed over 90\% of the total candidate pairs. If an unknown planet is close to the galactic plane or the galactic centre, it would not be identified in this work.

Zodiacal dust within the solar system may also be a cause of false positives, and will be considered below when reviewing images of shortlist candidates.

\subsection{Filters based on absolute and comparative flux measurements}

The next stage was to apply a series of filters to reduce the four lists of potential pairs down to a shortlist of potential candidates. The filters are intended to quickly eliminate pairs which are clearly not from a single source. The filters used evolved as large numbers of plots were viewed and common patterns recognised. The filters were deliberately set at fairly loose values in order not to make false exclusions, so our sample would retain a high level of completeness. This was particularly important since the flux measurements often included upper limits in the catalogues. The final set of flux filters used was:

(a) remove sources for which the fluxes were unrealistically high to be from potential planets; this cutoff was calculated for candidate mass $>$5 M$_{\rm J}$.  As an example, this gives a maximum of 9.22 Jy at 90 $\mu$m at 2000 AU;

(b) remove sources for which the {\it AKARI} and IRAS fluxes which were close in wavelength (60 - 65 $\mu$m and 90 - 100 $\mu$m) were strongly inconsistent. This was defined as a factor of 4, i.e. S$_{60}$/S$_{65}>4$; S$_{65}$/S$_{60}>4$; S$_{90}$/S$_{100}>4$; or S$_{100}$/S$_{90}>4$; 

(c) remove candidates with excessive slopes in one of the catalogues: S$_{90}$/S$_{65}>3$ or S$_{65}$/S$_{90}>3$ in {\it AKARI} or S$_{100}$/S$_{60}>3$ in IRAS;

(d) remove candidates with increasing {\it AKARI} fluxes at the high wavelengths: S$_{160}>$S$_{90}$; S$_{140}>$S$_{90}$;

(e) remove candidates with very different colours: the {\it AKARI} colour S$_{90}/$S$_{65}$ was compared with the IRAS colour S$_{100}/$S$_{60}$, and candidate pairs where the colours differed (in either direction) by a factor of over 3 were excluded; and

(f) remove candidates where the IRAS 25, 60 and 100 $\mu$m fluxes were all upper limits, and were monotonically increasing.

\begin{figure}
\begin{center}   
\resizebox{2.57in}{9.0in}{\includegraphics{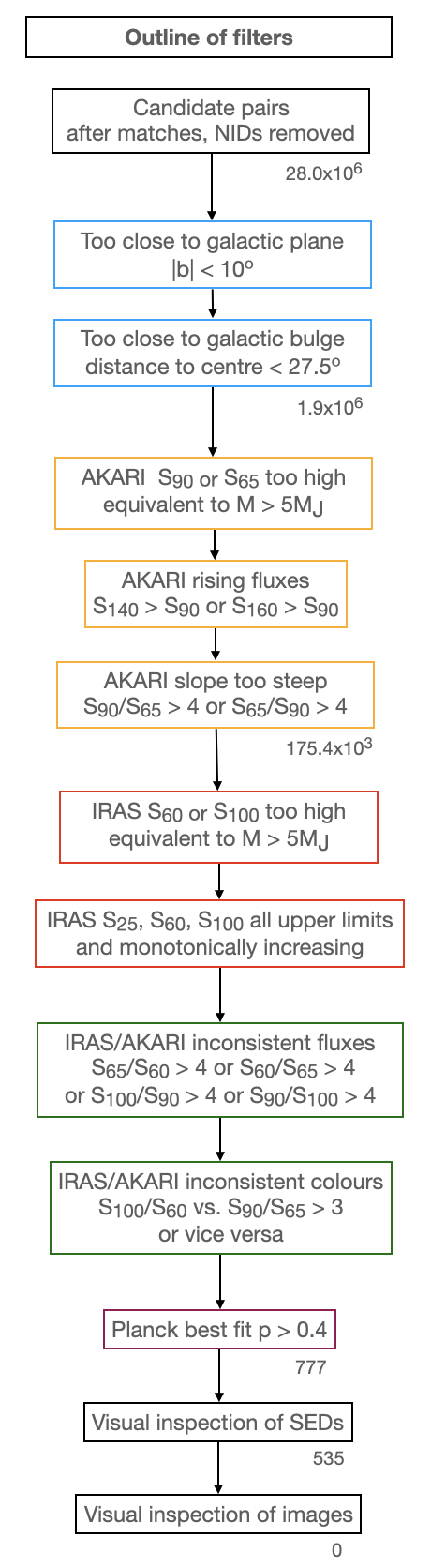}} 
\caption{Outline of the filters used to find candidate pairs. The numbers relate to the full search to date, from 8,000 AU to 700~AU.}\label{fig:filters}
\end{center}
\end{figure}

A schematic diagram of the filters is shown in Figure \ref{fig:filters}.

\subsection{Fitting a Planck blackbody distribution}\label{sec:fitting}

The next stage was to find the best-fit Planck distribution to seven of the eight {\it AKARI} and IRAS flux values using $\chi^2$ minimisation, for mass values in the range from 0.02 M$_{\rm J}$ (about half the mass of Uranus and Neptune) to 5~M$_{\rm J}$. 

For sources with M $>$ 1M$_{\rm J}$,  tables from Burrows et al.\ (1997) were used to estimate the radius and effective temperature for a given mass and age of a planet. For sources below 1M$_{\rm J}$, data from the four gas giants were used. The minimum value of $\chi^2$ was then calculated in the usual way. 

For the best-fit distribution, we calculated the probability that a value of $\chi^2$ would be this value or lower as:

 \begin{equation}
{\rm p}=1-{\rm pdf}(\chi^2_{\rm min},  ~{\rm dof})\label{equation:chi_pdf}
\end{equation}

\noindent where dof represents the number of degrees of freedom and the pdf is calculated with the Python function {\sc scipy.stats.chi2}.

In the small number of cases where SCANPI data was not available, we took the upper limit from PSCR as the value of the flux and estimated the error as 50\% of this value.

The Planck blackbody distribution is likely to be a fairly crude approximation to the detectible flux. In particular, emission from hydrocarbons dominates the mid-infrared emission of the gas giants. For Uranus, there is a pronounced peak between $\sim$12 $\mu$m and 15 $\mu$m: see Figure 3 in the recent review by Fletcher (2021). A similar peak from hydrocarbons in the near-infrared was found for Neptune by {\it AKARI} infrared spectroscopy (Fletcher et al. 2010). The thermal evolution models of Fortney et al.\ (2016) based on variations of Neptune-like parameters to model a potential Planet Nine also found flux emission up to $\sim$12 $\mu$m much higher than that expected from a blackbody spectrum at plausible temperatures. Therefore, we did not use the lowest-wavelength IRAS measurement (at 12 $\mu$m) in fitting the Planck distribution. Fletcher (2020) also found a  broad peak from H$_2$-He collisions  at $\sim$46-60 $\mu$m, the tail of which may affect the 60/65 $\mu$m values observed.

For flux values given as upper limits at other wavelengths, we have used the upper limit as the measurement and taken the error as 50\% of this value. We have also limited other errors to a maximum of 50\% of the measurement so that all the measurements (particularly the 140 $\mu$m and 160 $\mu$m {\it AKARI} measurements) are given some weight in the chi-square calculation, more than may be justified on a strictly statistical basis.

\subsection{Identify shortlist candidates}

Finally, we found a shortlist of planetary candidates by setting minimum $\chi^2$ probabilities. In view of the errors inherent in the data, and also the fact that planetary emission is likely to be more complicated than that given by a single Planck distribution, this has been essentially set to eliminate the sources whose best-fit Planck distributions are really poor fits. Setting p$>$0.4 yielded 777 sources, whose SEDs were then visually inspected. About one-third of these SEDs were clearly not viable candidates. Reasons for rejection included cases where most of the points were upper limits, and cases in which the IRAS and {\it AKARI} points did not look like they belonged to a single source. This left a final shortlist of 535 candidates, of which most (501) had IRAS observations from the IRAS-PSCR catalogue. To illustrate the results at this point, three of the best SED fits of these candidates are plotted in Figure \ref{fig:four_seds}. The overall results are summarised by AU range in Table~\ref{table:planets4}. 

\begin{figure}
\begin{center} 
\resizebox{3.0in}{!}{\includegraphics{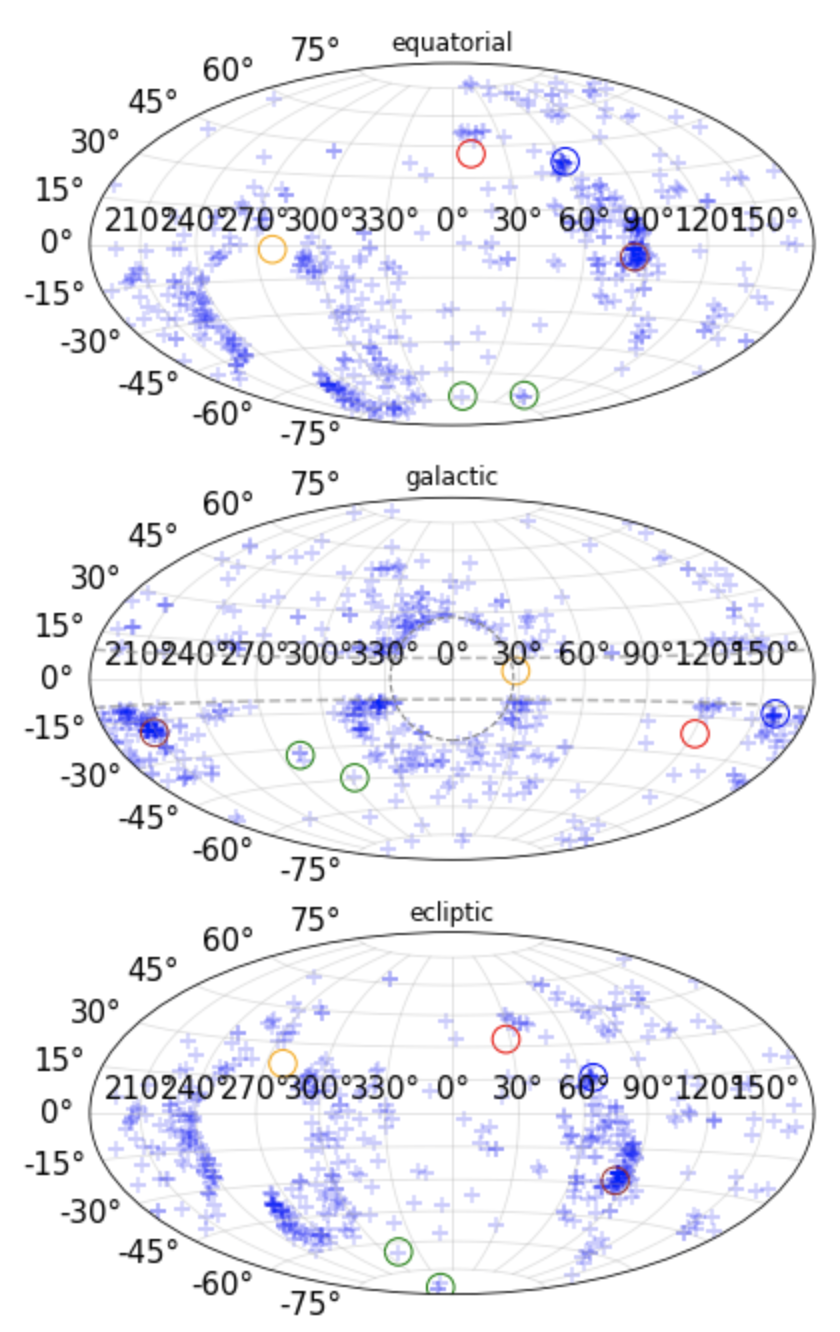}}  
\caption{The 535 shortlist candidates plotted as blue plus signs in equatorial, galactic and ecliptic coordinate systems. The $|b|<10^{\rm o}$ and galactic bulge exclusion zones  are shown in the galactic coordinates plot with grey dashed lines and a circle. Other regions which may be a source of cirrus are shown by circles: the magenta circle is centred on the position of Andromeda; the green circles on the Small and Large Magellanic Clouds; the brown circle is centred on the Orion nebula; the orange circle is the Gould strip; the blue circle on the California nebula, a nearby large HII region (circles are not to scale).}\label{fig:systems}
\end{center}
\end{figure}

\begin{figure*}
\begin{center}   
\resizebox{2.33in}{2.3in}{\includegraphics{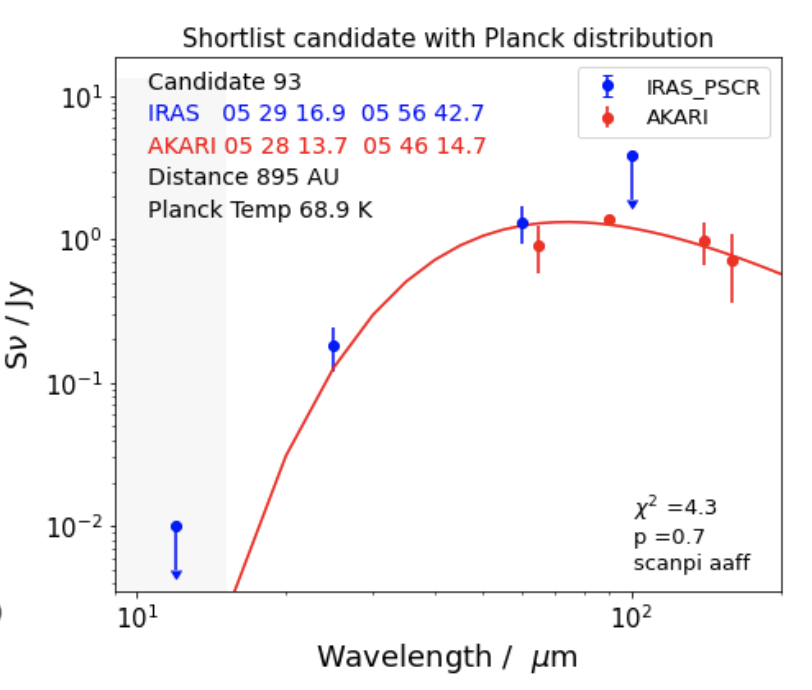}}
\resizebox{2.3in}{2.3in}{\includegraphics{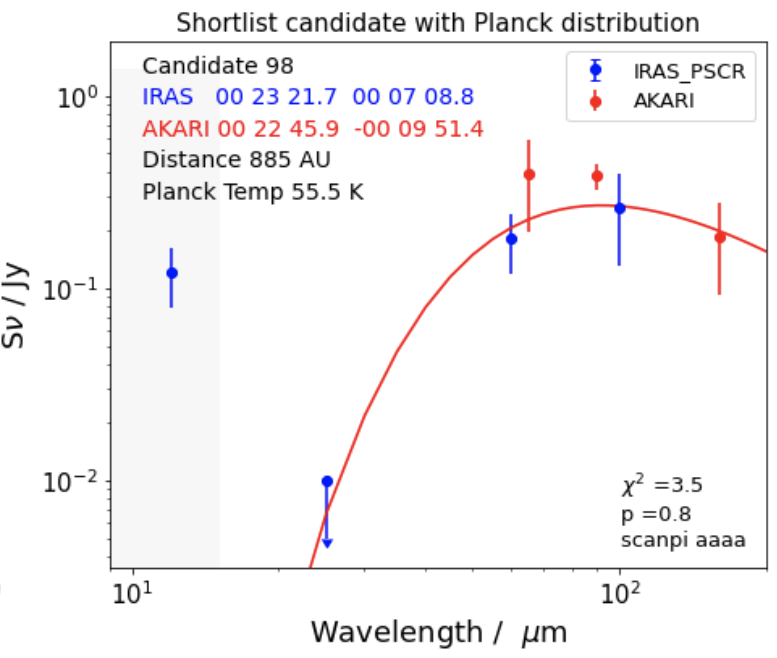}} 
\resizebox{2.3in}{2.3in}{\includegraphics{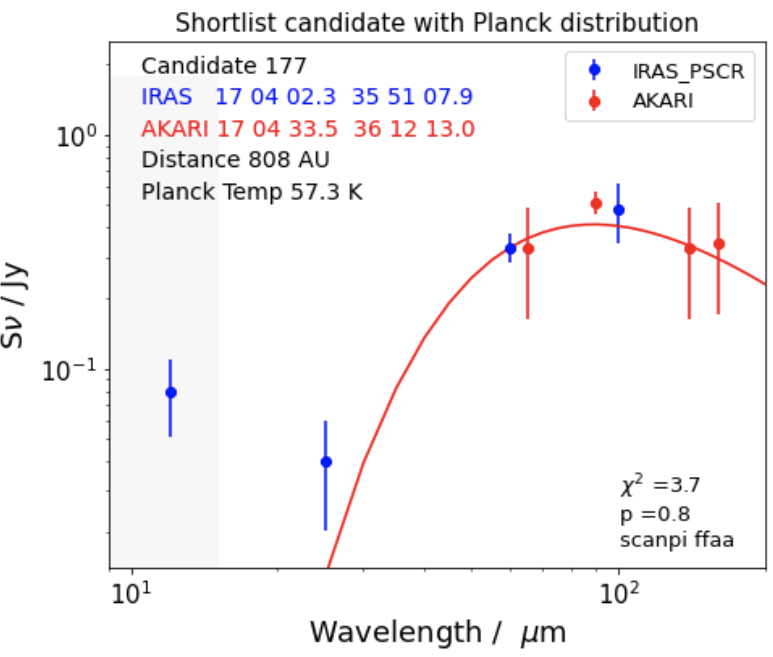}} 
\caption{ Examples of plausible SED fits:  flux measurements and SEDs of the best-fit Planck distribution for three of the shortlist candidates. Note that the 12~$\mu$m IRAS points are not included in the minimum-$\chi^2$ fitting. The grey band relates to expected hydrocarbon emission between 6-15~$\mu$m. The bottom right of each figure shows the SCANPI flux used (a=amplitude, p=peak, f=average between fixed points).}\label{fig:four_seds}
\end{center}
\end{figure*}

Figure \ref{fig:systems} shows a plot of the shortlist candidates on the sky in three celestial coordinate systems and reveals several \textquoteleft clumps\textquoteright~of sources close to the galactic plane cutoff and the galactic bulge exclusion circle. Some sources  coincide with known objects: in particular, there are sources  which coincide with a nearby large HII region called the California Nebula, and a clump of sources are in the Orion Nebula. There were no shortlist candidates near the positions of Andromeda, M82 or Arp220.
 
We did not find any candidates exceeding the mass of Jupiter, and as our search reached 1,000 AU and below, most candidates were around Neptune/Uranus mass. It is interesting that the most common exoplanets in the most recent census have sub-Neptune masses (Fulton \&  Petigura 2018).


\begin{table}
\caption{Number of candidates in AU ranges showing the total pairs, and details of the final shortlist after all the filters (including Planck $\chi^2$ best-fits with p$>0.4$ and visual inspection of SED) but before any images were examined.}\label{table:planets4}
\begin{center}
\begin{tabular}{|lrrrrr|}
\hline   

Distance               & Separation             & Total                       &   \multicolumn{3}{c|}{------ Shortlist ------}          \\
range                    &   range                   &  pairs                      &                          &  \multicolumn{2}{c|}{mass range }      \\  
(AU)                       & (arcsec)                &    (10$^3$)                           &           (no.)             & (M$_{\rm J}$) & (M$_{\rm J}$)\\          

 \hline
 
6000 - 8000         & 42 - ~~~~65           &   52                      &     0                    &                                 \\
4000 - 6000         & 65 -~~ 120             &  132                      &    1                   &              &    0.36        \\         
2000 - 4000         & 120 -~~ 339           &  1,065                 &      6                   &   0.13      & 0.25        \\     

1500 - 2000             & 339 -~~ 523        &  1,688                 &    12                  & 0.09          &   0.16      \\   
1000 - 1500              &  523 -~~ 960      &  6,812                 &    42                  &  0.05         & 0.18        \\   
~~900 - 1000             & 960 - 1124        &   3,572                 &    31                  &  0.03       & 0.08         \\  
~~800 -~~900           &  1124 - 1341      &  5,561                  &   90                  &  0.02       &   0.11        \\  
~~750 -~~800          & 1341 - 1478        & 3,979                 &   137                  &  0.02        &  0.17      \\
~~700 -~~750          & 1478 - 1639        & 5,183                 &   216                  &  0.02        &  0.14      \\\\  

Total                   &                              &     28,044               &   535                   & 0.02      &   0.36      \\     
\hline
\end{tabular}
\end{center}
\end{table}

\section{Examining the shortlist candidates}\label{sec:shortlist}

Next, image cutouts were downloaded from publicly available sites for IRAS and \textit{AKARI}, giving images at 8 wavelengths for each of the 535 candidates. 

None of the {\it AKARI} images showed a source at the expected location. For IRAS, sources close to the position of candidates showed in the images in seven cases, but all were in clouds of cirrus and were not convincing identifications of potential planets. 

The images proved to be fairly conclusive: not only was no evidence seen of the candidates in the images, but the images for IRAS 60 and 100 $\mu$m and for AKARI 90 $\mu$m and the higher wavelengths showed clear evidence of cirrus for almost all 535 sources. Figure~\ref{fig:cirrus} shows some examples of this. Clearly the flux is from the cirrus, which peaks in the far infrared, and not from a planet.
 
There were only four candidates whose location in IRAS images showed no cirrus, and two cases where our candidate positions appeared to be beyond the edge of nearby cirrus. The four without cirrus are all close to the Orion nebula (see Figure \ref{fig:systems}). Images and SEDs for two of these are shown in Figure \ref{fig:cirrus}. The flux for the two candidates outside the edge of the cirrus cloud is also likely to be from cirrus, since the boundaries of the clouds are not well-defined.

We  therefore conclude that we have not found a new planet in the region we have explored to date.

If any of the candidates had survived at this point, we would need to check whether any of the candidates is already a known source - the matching to catalogues done by IRAS which we used at the start of the work was from nearly forty years ago. We would also check whether either member of the pairs shows up on catalogues of asteroids etc. as discussed above. 
 
If the candidates had survived these checks, we would use the multiple IRAS scans to check for evidence of parallax, and we would extrapolate the locations to check for confirmation from WISE catalogues and images.
 We estimate the positional error for follow-up observations would be about 25$\arcsec$, considering the positional accuracy of the two catalogues, proper motion error and the parallax for six months.

\section{Discussion}\label{sec:results}

The method we have used should have a good possibility of identifying unknown planets in the solar system, if they exist and are not currently located in cirrus, as we work inwards from 700 AU, covering more of the region identified in Batygin review mentioned earlier. We will also reduce the lower bound for the planet mass as we move closer and this becomes observable with these surveys. It is worth noting that we have only just reached the flux detectability limit for our surveys for a Uranus-mass planet at 700 AU (see Table \ref{table:proper_motion}).

 The main weakness in the method we have used (apart from being dominated by cirrus) has been the limited data available for fitting with an SED. The IRAS 12 $\mu$m was not used, as discussed above, reducing the number of data points from 8 to 7. The AKARI 140 and 160~$\mu$m points were sometimes missing, further reducing the number of points to five. In addition, there were many cases of fluxes being upper limits. Also, of course, a planet could reside near the galactic centre, or within a cloud of cirrus, and be missed by this work.
 
A significant proportion of the emission from interstellar dust occurs in the far-infrared region of the spectrum which we are using in this work, so it is not surprising that the locations of so many of our putative candidates have turned out to be regions of dusty material. Investigation of dusty material in the ISM was indeed a major objective of the {\it AKARI} survey maps (Doi et al.\ 2017).
 
 Nevertheless, if there is an unobscured planet away from the galactic plane and centre at several hundred AU, this method has a good chance of finding it, and we are planning to extend inwards from 700 AU in the near future. One advantage of this method is that, if a good candidate for a planet is detected, we would have a good fix on its coordinates for follow-up confirmation.

\section{Acknowledgements}
This work has used observations from IRAS and from {\it AKARI}, a JAXA project with participation of ESA. We have used the data archives and tools provided by IRSA. We would like to thank the referee, Michael Rowan-Robinson, for his constructive comments on our original draft, which led to significant improvements in the work.

\section{Data Availability}
All data used in this paper have been taken from public archives.


\begin{figure*}
\begin{center} 
\resizebox{1.72in}{!}{\includegraphics{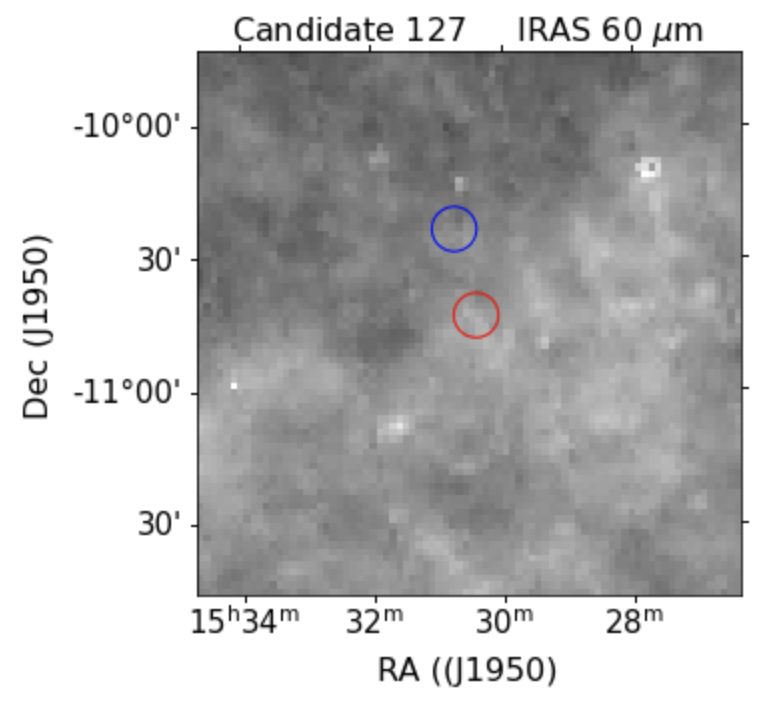}} 
\resizebox{1.70in}{!}{\includegraphics{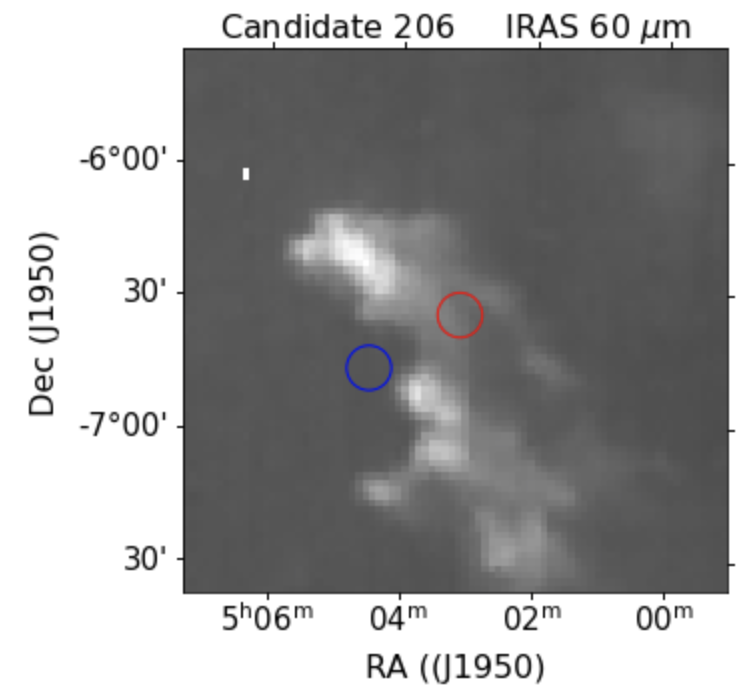}} 
\resizebox{1.70in}{!}{\includegraphics{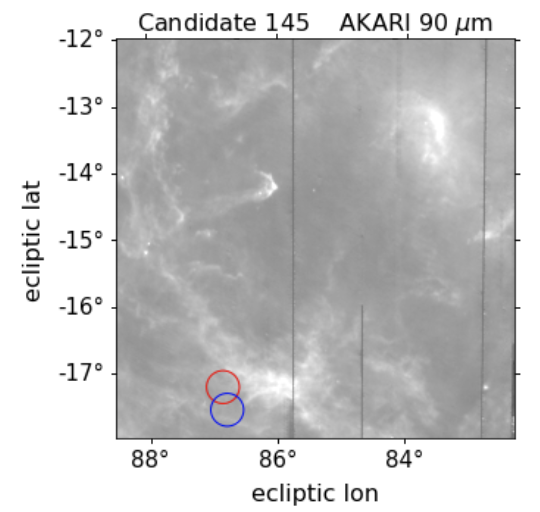}} 
\resizebox{1.63in}{!}{\includegraphics{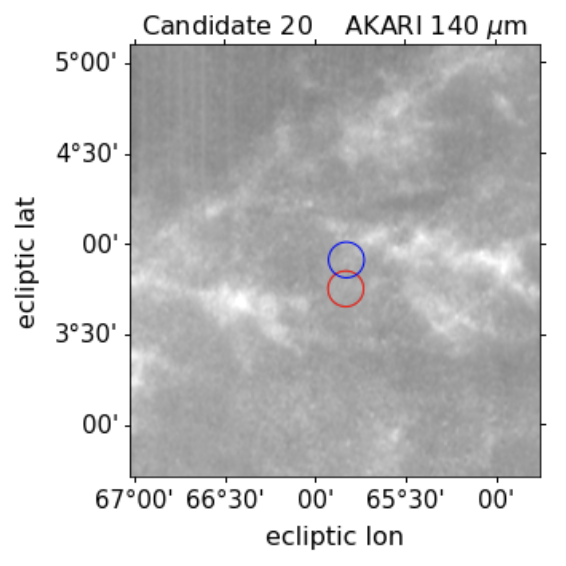}} 
\caption{Examples of IRAS and {\it AKARI} images showing candidates (like almost all of the shortlist candidates) inside clouds of cirrus which is almost certainly the cause of the far-infrared flux. Circles indicate target coordinates: red for IRAS; blue for AKARI. The size of circles is arbitrary. Note IRAS images are in J1950 equatorial coordinates; AKARI in ecliptic coordinates.}\label{fig:cirrus}
\end{center}
\end{figure*}

\begin{figure*}
\begin{center} 

\resizebox{2.5in}{2.3in}{\includegraphics{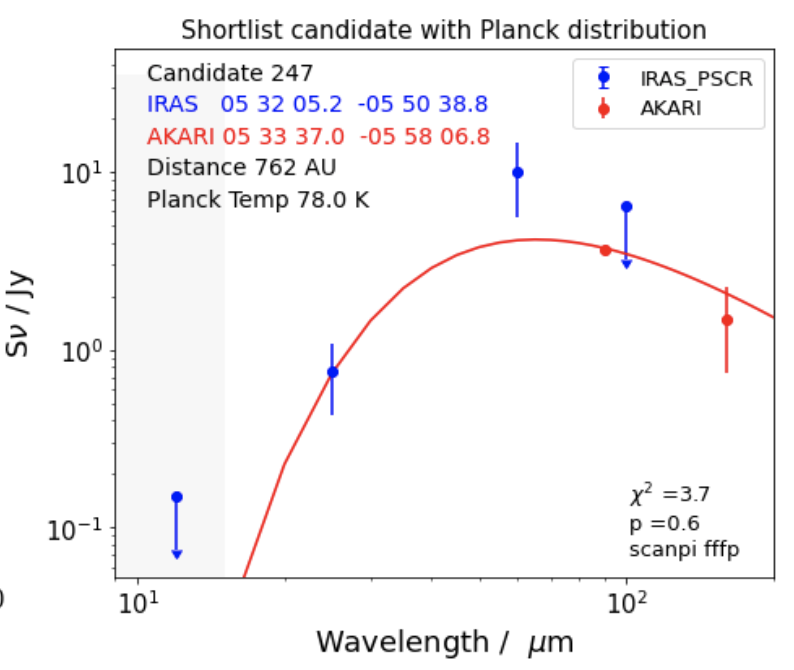}} 
\resizebox{2.5in}{2.31in}{\includegraphics{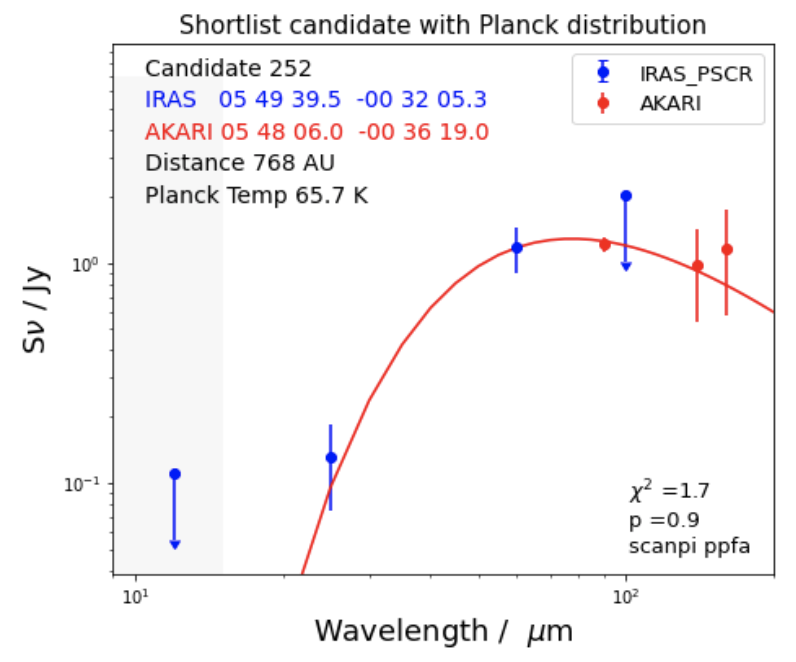}} 
\resizebox{2.2in}{!}{\includegraphics{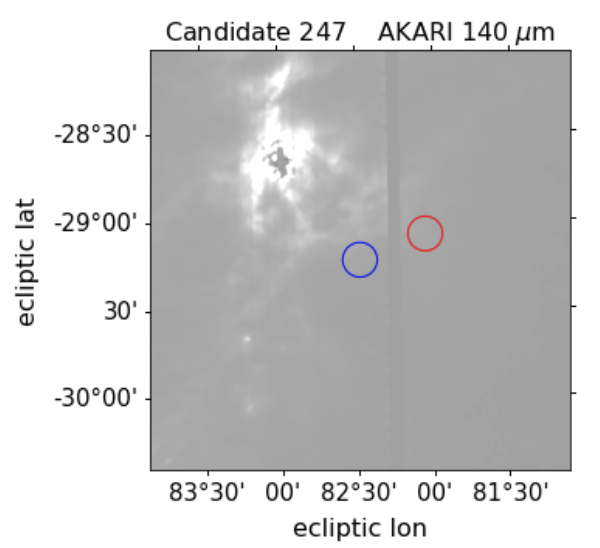}} 
\resizebox{2.25in}{!}{\includegraphics{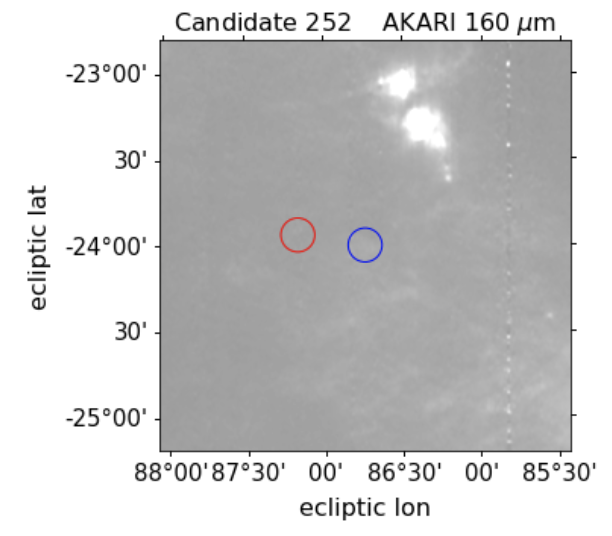}} 
\caption{ Although we have concluded these are not planets, we show above the results for two of our best shortlist candidates. Target coordinates are located outside cirrus clouds in some of their images, but  higher-wavelength {\it AKARI} images seem to suggest that their flux, like that of other candidates,  probably comes from outlying cirrus. Circles indicate target coordinates: red for IRAS; blue for {\it AKARI}. The size of circles is arbitrary. Note IRAS images are in J1950 equatorial coordinates; {\it AKARI} images are in ecliptic coordinates.}\label{fig:cirrus}
\end{center}
\end{figure*}


\bsp   
\label{lastpage}
\end{document}